\documentclass[aps,twocolumn,pra,superscriptaddress]{revtex4}
\usepackage{epsfig,graphicx,times}
\usepackage{amstext}
\usepackage{amsmath}
\usepackage{lipsum}
\usepackage{amssymb}
\usepackage{float}
\usepackage{graphicx}
\usepackage{latexsym}
\usepackage{bm}
\usepackage{epstopdf}
\usepackage{appendix}
\usepackage[none]{hyphenat}
\usepackage[colorlinks,citecolor=blue, linkcolor=blue,hyperindex,CJKbookmarks]{hyperref}
\newcommand{\Tr}{\text{Tr}}

\begin{document}

\title{Quantum sensing of aging transitions}

\author{Huining Zhang}
\affiliation{Center for Quantum Sciences and School of Physics, Northeast Normal University, Changchun 130024, China}
\author{Yunbo Zhang}
\affiliation{Zhejiang Key Laboratory of Quantum State Control and Optical Field Manipulation, Department of Physics, Zhejiang Sci-Tech University, Hangzhou 310018,  China}
\author{Xiaoguang Wang}
\affiliation{Zhejiang Key Laboratory of Quantum State Control and Optical Field Manipulation, Department of Physics, Zhejiang Sci-Tech University, Hangzhou 310018, China}
\author{X. X. Yi\footnote{yixx@nenu.edu.cn}}
\affiliation{Center for Quantum Sciences and School of Physics, Northeast Normal University, Changchun 130024, China}
\affiliation{Center for Advanced Optoelectronic Functional Materials Research, and Key Laboratory for UV Light-Emitting Materials and Technology of Ministry of Education, Northeast Normal University, Changchun 130024, China}

\date{\today}

\begin{abstract}
The aging transition is a critical phenomenon in which collective dynamics deteriorate as the fraction of inactive quantum nodes exceeds a threshold, referred to as the aging transition point. Such transitions are relevant to a broad range of biological and physiological systems, and may play an important role in quantum information processing, particularly in the stability assessment and robustness control of quantum networks. Detecting the aging transition point is therefore crucial for predicting network breakdown, since it marks the critical threshold at which a quantum network abruptly loses its stable active state and enters a degraded inactive phase.
Here we propose a quantum sensing strategy to locate this transition point using a single qubit probe coherently coupled to a small subset of oscillator nodes. As the inactive fraction $p$ approaches the aging transition point, the excited-state population of the probe becomes highly sensitive to variations in $p$, leading to a pronounced enhancement of the Fisher information. This critical enhancement enables high-precision estimation of the transition point. Remarkably, this enhancement survives even in the classical regime for the oscillators, where the Fisher information increases dramatically as $p$ approaches the transition region. Our results establish a feasible route to sensing aging transitions in oscillator networks and provide a metrological perspective on critical phenomena in quantum many-body systems.
\end{abstract}
\maketitle

\section{INTRODUCTION}

Coupled oscillators provide a versatile platform for exploring many-body physics, with the aging transition serving as a paradigmatic example of collective dynamical criticality \cite{Daido2004,Bandy2023,Daido2007,Daido2008,Sun2019,Sun2017,Sath2019,Sath2022,Ponrasu2020,Biswas2022,Sahoo2023,Singh2020,Rahman2017,Rakshit2020,Zou2021,Zhang2025}. In biological and physiological systems, aging is associated with the progressive degradation of functional activity essential for survival and regenerative capacity. In oscillator networks, aging occurs when a fraction of oscillators becomes inactive \cite{Daido2004}. For classical oscillators, active and inactive nodes correspond to self-oscillatory and non-self-oscillatory units, respectively. As the fraction of inactive oscillators increases, the network undergoes an aging transition, beyond which collective oscillations cease \cite{Daido2004}. In quantum oscillator networks \cite{Bandy2023,Zhang2025}, activity is characterized by dissipative processes: single-boson gain defines an active quantum oscillator, whereas single-boson loss defines an inactive one. As the inactive fraction increases, the mean boson number develops a critical knee point separating two distinct decay regimes, thereby signaling the aging transition in coupled quantum oscillators.

Identifying the aging transition point is essential for predicting and mitigating the breakdown of collective \mbox{oscillations} in coupled-oscillator networks. Its precise detection is also crucial for ensuring the stable operation of practical systems, including transportation networks \cite{Ge2004}, power grids \cite{Rohden2012}, biological organisms \cite{Cama2003}, and generic physical systems \cite{Hopf1982}. Nevertheless, high-sensitivity estimation of the aging transition point remains an open challenge. Here we address this problem by leveraging quantum sensing techniques.

Quantum sensing \cite{Deg2017} exploits  quantum features such as coherence and entanglement to estimate  physical \mbox{parameters}. In recent years, this concept has developed into a rapidly expanding field of quantum science and technology. Many critical behaviors, such as dissipative phase transitions, have attracted considerable attention in designing quantum sensing protocols  \cite{Rag2018,Heu2019,Fer2017,Ili2022,Can2023,Chu2021,Lu2001,
Iva2013,Sal2014,Bin2016,Ram2018,Wal2020,Fre2018,Gar2020,Nie2021,
Gar2022,Che2024,Liu2021,Din2022} that can harness criticality to enable precise parameter estimation. The  physical realizations of quantum sensors include ultracold atoms \cite{Jes2001,Rosi2013}, trapped ions \cite{Wine2005,Tim2008}, superconducting qubits \cite{Reed2012,Sporl2007} and  electronic spin qubits \cite{Wrach2006,Wald2014}. These systems  hold promise for achieving high sensitivity and precision in a variety of applied and fundamental settings.

In this work, we apply quantum sensing to estimate the aging transition in a quantum oscillator network.
Detecting such a transition typically requires global measurements or full knowledge of the oscillator ensemble. However, from a practical  perspective, it is highly desirable to use a local probe that can infer the onset of aging without disturbing the critical behavior of the network. Here, we propose and analyze a quantum sensing protocol where a single qubit is coherently coupled to  a finite small group (active, inactive, or both) of the oscillators. The qubit acts as a sensitive probe, capable of precisely sensing the aging transition point. Its excited-state population responds sharply to changes in the inactive fraction $p$ as $p$ increases toward the aging transition point. And the Fisher information exhibits a critical enhancement, confirming the potential for quantum metrology. Moreover, we demonstrate that this critical enhancement persists even in the classical limit of large occupation numbers, highlighting the robustness of the sensing scheme.

This work is organized as follows. In sec.~\ref{sec1}, we present  a model consisting of active and inactive quantum oscillators. A dissipative qubit coupled to several oscillators is introduced to serve as the quantum probe. We write a full quantum master equation to describe  the coupled dynamics of the oscillator network   and the probe qubit.  In Sec.~\ref{sec2}, we demonstrate  the performance of the qubit probe to estimate  the aging transition point of the quantum oscillator system, with the probe coupled to one, two, and three active oscillators respectively. A detailed analysis of the results is also presented.  In Sec.~\ref{sec3}, we extend our investigation to the classical limit, and explore the capability of the qubit probe to estimate the aging transition of the oscillator system in this regime. In Sec.~\ref{sec4}, we analyze the quantum Fisher information. Finally, we summarize the results of this work in Sec.~\ref{sec5}. Appendix~\ref{appendixA} gives the specific forms of the mean-field equations based on  the master equation. Details for the the nonlinear matrix $\boldsymbol{M}$ are given in Appendix~\ref{appendixB}.

\section{MODEL}\label{sec1}
\begin{figure}[t]
	\centering
	\includegraphics[width=0.45\textwidth]{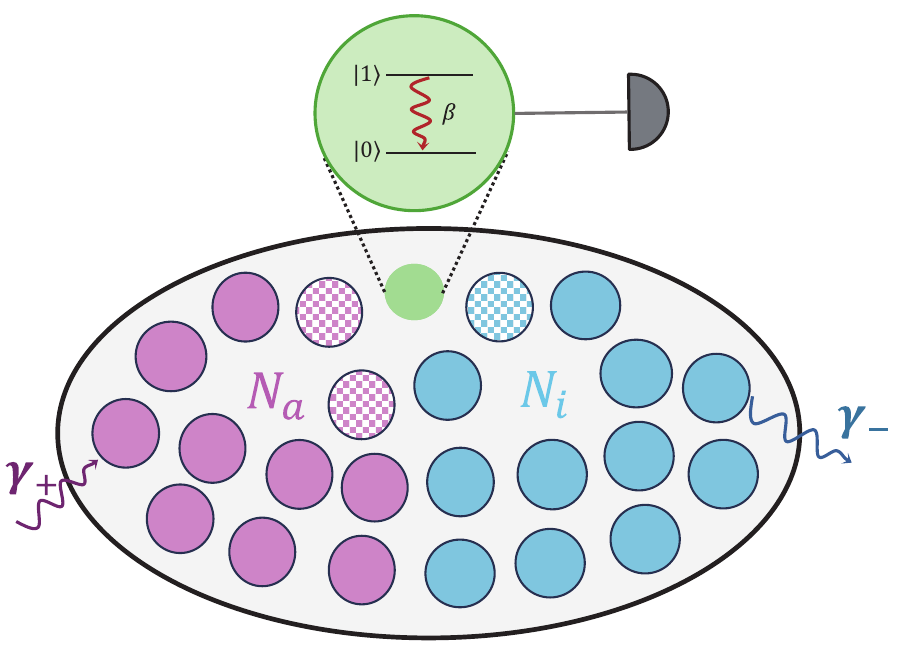}
	\caption {The quantum system consists of $N_a=N(1-p)$ active oscillators with gain rate $\gamma_+$ and $N_i=Np$ inactive oscillators with loss rate $\gamma_-$. A dissipative qubit, with decay rate $\beta$, serves as a local probe for detecting the aging transition point of the oscillator network. The checkerboard lattice denotes the subset of oscillators coherently coupled to the qubit for quantum sensing.}
	\label{fig1}
\end{figure}
The system considered in this work comprises $N$ quantum oscillators with globally dissipative coupling. We suppose that $N_a$ oscillators (with $N_a\leq N$)  are active, while the remainders  $N_i=N-N_a$  are inactive. The ratio of inactive to total oscillators is defined by \(p = \frac{N_i}{N} = \frac{N - N_a}{N}\). There exists a critical ``knee'' value  in the fraction of inactive oscillators around which quantum aging occurs in two different ways. This knee point is referred to  aging transition point $p_c$ \cite{Bandy2023,Zhang2025}. The aging transition points increase with dissipative coupling for the quantum oscillators, whereas they decrease for the classical oscillators \cite{Daido2004,Bandy2023}.

In this work, a qubit is introduced and coherently coupled to part of the oscillators, acting as a probe to detect the aging transition point $p_c$. This proposal is illustrated in  Fig.~\ref{fig1}. The Hamiltonian of such a system  in the rotating wave approximation is given by (with $\hbar=1$)
\begin{equation}
H=\frac{1}{2}\omega_0\sigma_z+\sum_{k=1}^N\omega_k a_{k}^{\dag}a_k+g\sum_{k\in S}(a_{k}^{\dag}\sigma_{-}+a_{k}\sigma_{+}).
\label{Hamiltonian1}
\end{equation}
Here $\omega_0$ and $\omega_k$ denote the transition frequency of the qubit and the frequency of the $k$th oscillator, respectively. The operators $a_k^\dagger$ and $a_k$ are the creation and annihilation operators of the $k$th oscillator. The Pauli operators are defined as $\sigma_z=|1\rangle\langle 1|-|0\rangle\langle 0|$, $\sigma_+=|1\rangle\langle 0|$, and $\sigma_-=|0\rangle\langle 1|$, where $|1\rangle=(1,0)^T$ and $|0\rangle=(0,1)^T$ denote the excited and ground states of the qubit, respectively. The first two terms in Eq.~(\ref{Hamiltonian1}) describe the free Hamiltonians of the qubit and the oscillator network, while the third term describes coherent coupling between the qubit and a subset $S$ of oscillators, with coupling strength $g$. The subset $S$ may contain active oscillators, inactive oscillators, or both.

In order to address the problem easily, we transform the system into the interaction picture by defining $U=\text{exp}[-i(\frac{1}{2}\omega_0\sigma_z+\sum_{k=1}^N\omega_k a_{k}^{\dag}a_k)t]$, the Hamiltonian is then
\begin{equation}
\begin{aligned}
H_1&=g\sum_{k\in S}(a_{k}^{\dag}\sigma_{-}e^{-i\Delta_k t}+a_{k}\sigma_{+}e^{i\Delta_k t}),
\label{Hamiltonian2}
\end{aligned}
\end{equation}
where $\Delta_k=\omega_0-\omega_k$. In what follows, we assume that all oscillators in $S$ are degenerate and resonantly coupled to the qubit, i.e., $\Delta_k=0$.

The dynamics of $N$ quantum oscillators with diffusive global coupling and a dissipative qubit is governed by the quantum master equation \cite{Bandy2023,Zhang2025,Lee2014,Ishi2017}
\begin{widetext}
\begin{equation}
\dot{\rho}=-i[H_1,\rho]+\sum_{k=1}^{N_a} \gamma_{+} \mathcal{D}[a_{k}^{\dag}](\rho) + \sum_{k=N_a+1}^{N} \gamma_{-} \mathcal{D}[a_{k}](\rho)+\sum_{k=1}^{N}\kappa\mathcal{D}[a_{k}^2](\rho)+
{\sum_{j,k}}^{'}\frac{V}{N}\mathcal{D}[a_{j}-a_{k}](\rho)+\beta\mathcal{D}[\sigma_{-}](\rho).
\label{master equation}
\end{equation}
\end{widetext}
Here $\rho$ is the density matrix of the full system, and $\mathcal{D}[\hat{\mathcal O}](\rho) = \hat{\mathcal O}\rho\hat{\mathcal O}^{\dagger} - \frac{1}{2}\{\hat{\mathcal O}^{\dagger}\hat{\mathcal O},\rho\}$ denotes the Lindblad dissipator. The dissipative channels $\mathcal D[a_k^\dagger]$ and $\mathcal D[a_k]$ distinguish active and inactive oscillators, respectively \cite{Bandy2023,Zhang2025}. Active oscillators, labeled by $k\in \mathcal N_a={1,\ldots,N(1-p)}$, undergo single-boson gain at rate $\gamma_+$, whereas inactive oscillators, labeled by $k\in \mathcal N_i={N(1-p)+1,\ldots,N}$, undergo single-boson loss at rate $\gamma_-$. This notation follows Refs.~\cite{Bandy2023,Zhang2025}. The parameter $\kappa$ denotes the nonlinear damping rate, while $V$ characterizes the dissipative coupling between distinct oscillators; the prime on $\sum_{j,k}'$ excludes terms with $j=k$. The coefficient $\beta$ is the qubit decay rate.

The model combines local gain or loss, nonlinear damping, and dissipative global coupling, which together generate a nonequilibrium competition between activity and dissipation. Increasing the inactive fraction $p$ shifts this balance and eventually suppresses collective oscillations, giving rise to the aging transition. The coherently coupled dissipative qubit serves as a local probe of this collective instability, allowing the transition point to be inferred without global measurements of the oscillator network.

\section{Detection of the aging transition point in the  quantum regime}\label{sec2}

For a large number of particles,  the mean-field approximation allows the many-body density matrix to be factorized as \cite{Dieh2010}
\begin{equation}
\begin{aligned}
\rho(t) \approx \rho_a(t) \otimes \bigotimes_{k=1}^{N} \rho_k(t),
\label{equation4}
\end{aligned}
\end{equation}
where $\rho_a$ and $\rho_k$ are the one-body density matrices of the qubit and the $k$th ($k=1,\dots,N$) oscillator, respectively.

Assuming identical oscillators within each group, we divide the network into four groups. Group $A$ consists of active oscillators that are not coupled to the qubit, all described by the same density matrix $\rho_A$; Group $I $ consists of inactive oscillators that do not interact with the qubit, all by the same density matrix $\rho_I$; Group $B_+$ consists of active oscillators that interact with the qubit described by the same density matrix $\rho_{B_+}$; and Group $B_-$ consists of inactive oscillators that interact with the qubit by the same density matrix $\rho_{B_-}$.
With this consideration together with the normalization and trace-preservation conditions, $\mathrm{Tr}\rho_k=1$ and $\mathrm{Tr}\dot{\rho}_k=0$, Eq.~(\ref{master equation}) reduces to a set of master equations of the form
\begin{equation}
\begin{aligned}
\dot{\rho_x}&=-i[H_x,\rho_x]+\mathcal{L}_x(\rho_x),
\label{equation5}
\end{aligned}
\end{equation}
where $x=a,A,I,B_+,B_-$ labels the corresponding master equations for the qubit and the oscillator groups, respectively. The details  of $H_x$ and $\mathcal{L}_x$ can be found in Appendix~\ref{appendixA}.

In the equation for $\dot{\rho}_{B_{\pm}}$, the operator $\hat O=a^\dagger$ $(a)$ denotes the gain (loss) channel of active (inactive) oscillators coupled to the qubit with rate $\gamma_+$ $(\gamma_-)$, respectively. Let $m$ active and $n$ inactive oscillators interact with the qubit. The mean-field term entering $\dot{\rho}_a$ is
$M=\sum_{j\in S}\mathrm{Tr}(a\rho_j)$,
which represents the sum of the expectation values of the annihilation operator $a$ over the one-body density matrices $\rho_j$ with $j\in S$. For the oscillator groups $x=A,I,B_+,B_-$, the corresponding mean-field terms take the form
$M_x=\sum_{j=1}^{N}{}^{\prime}\mathrm{Tr}(a\rho_j)$.
For the $k$th oscillator belonging to group $x$, the prime indicates that the   term with $j=k$ is excluded from the sum. The explicit expressions for $M_x$ are given in Appendix~\ref{appendixA}.

The mean-field approximation leads to a closed set of nonlinear master equations, given in Eq.~(\ref{equation5}). Since the quantum oscillators are bosonic, their Hilbert spaces must be truncated for numerical simulations. In the quantum regime, we focus on the parameter regime $\kappa>\gamma_\pm$, where strong nonlinear damping suppresses the occupation of highly excited Fock states and confines the dynamics close to the ground \mbox{state} \cite{Lee2013}. We therefore truncate the local Hilbert space at the boson number $n_0=4$, as higher Fock states are rapidly depleted by damping.

The sensor consists of a single qubit that can be coupled to a subset of, or all, oscillators. The population of the qubit excited-state, which serves as an indicator for detecting the aging transition point of the quantum oscillator network, is defined as
\begin{equation}
\begin{aligned}
P_e =\text{Tr}(|1\rangle\langle 1|\rho_a)=\text{Tr}(\sigma_+\sigma_-\rho_a).
\end{aligned}
\end{equation}
The excited-state population $P_e$ is obtained from the steady-state density matrix $\rho_a$ by solving Eq.~(\ref{equation5}) numerically. In our sensing protocol, $P_e$ is measured as a function of the inactive fraction $p$. By gradually increasing $p$, we monitor the response of the qubit as the oscillator network approaches the aging transition. We focus on the experimentally relevant case in which the qubit is coupled only to a small number of oscillators, which may be active, inactive, or a mixture of both. Since such local coupling only weakly perturbs the oscillator network, the position of the aging transition point remains essentially unchanged, allowing the transition to be accurately inferred from the qubit response. We first consider the qubit coupled to one, two, or three active oscillators. For example, when the probe is coupled to a single active oscillator, labeled as oscillator 1, the Hamiltonian in Eq.~(\ref{master equation}) reduces to
\begin{equation}
\begin{aligned}
H_1&=g(a_{1}^{\dag}\sigma_{-}+a_{1}\sigma_{+}).
\label{equation7}
\end{aligned}
\end{equation}
\begin{figure*}[t]
\centering
\includegraphics[width=8.6cm]{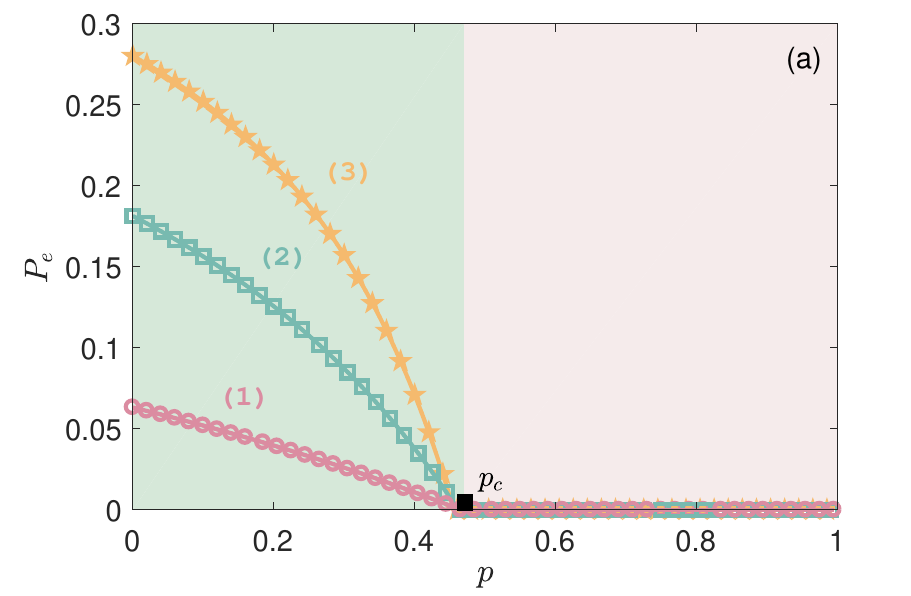}
\includegraphics[width=8.6cm]{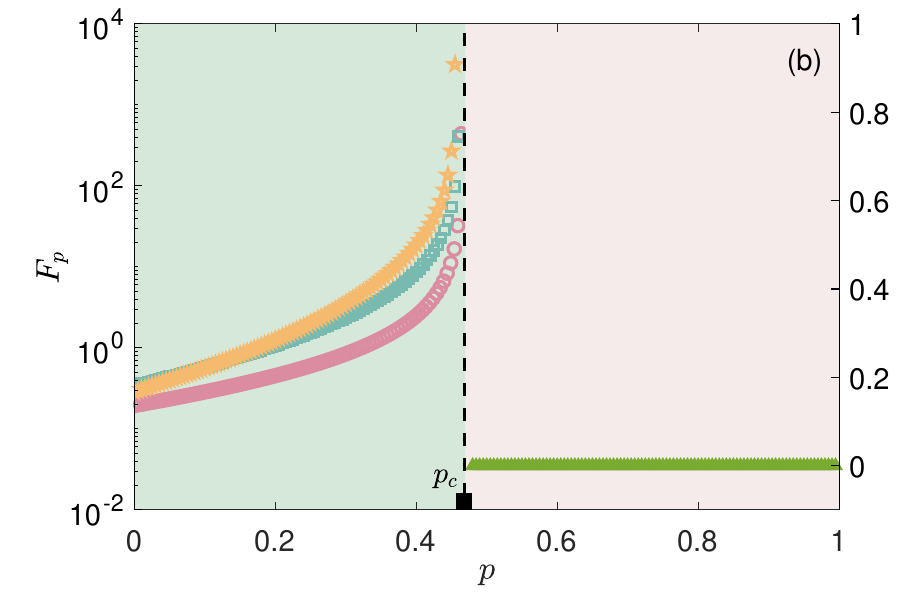}
\caption{Sensing performance in the quantum regime.
(a) Qubit excited-state population $P_e$ as a function of the inactive fraction $p$. The pink, green, and yellow curves correspond to the qubit coupled to one, two, and three active quantum oscillators, respectively.
(b) Fisher information $F_p$ for estimating the inactive fraction $p$. The green and pink regions indicate regimes with nonzero and zero $P_e$, respectively. In the green region, $F_p$ is plotted on a logarithmic scale, whereas in the pink region it is plotted on a linear scale. We set $\gamma_+=0.5\kappa$ for active oscillators and $\gamma_-=0.375\kappa$ for inactive oscillators. Other parameters are $N=200$, $\kappa=1$, $g=0.125\kappa$, $\beta=0.375\kappa$, and $V=5\kappa$.}
\label{aqbpic23}
\end{figure*}
The master equations in Eq.~(\ref{equation5}) then consist of $\dot{\rho}_a$, $\dot{\rho}_A$, $\dot{\rho}_I$, and $\dot{\rho}_{B+}$. Substituting $m=1$ and $n=0$ into Eq.~(\ref{equationA1}), we obtain the corresponding mean-field terms,
\begin{equation}
\begin{aligned}
M&=\text{Tr}(a\rho_{B_{+}}),\\
M_A&=
\sum_{j\in A}{}^{'}\text{Tr}(a\rho_A)+\sum_{j\in I}\text{Tr}(a\rho_I)+\text{Tr}(a\rho_{B_+}),\\
M_I&=
\sum_{j\in A}\text{Tr}(a\rho_A)+\sum_{j\in I}{}^{'}\text{Tr}(a\rho_I)+\text{Tr}(a\rho_{B_+}),\\
M_{B_{+}}&=
\sum_{j\in A}\text{Tr}(a\rho_A)+\sum_{j\in I}\text{Tr}(a\rho_I).
\label{equation8}
\end{aligned}
\end{equation}
The same applies to other situations for different $m$ and $n$.

In the present work, we propose to infer the aging transition  from measurements of the qubit excited-state population. To this end, we first characterize the sensing performance using the classical Fisher information\cite{Hel1976,Hol1982} associated with this specific measurement. This quantity is measurement dependent and quantifies the amount of information about the inactive fraction $p$ that can be extracted from the population measurement. It should therefore be distinguished from the quantum Fisher information, which gives the ultimate precision bound optimized over all possible measurements on the probe state. Comparing these two quantities allows us to assess how efficiently the population measurement exploits the parameter sensitivity encoded in the qubit, as we will discuss later. For the binary outcomes $|1\rangle$ and $|0\rangle$, with probabilities $P_e$ and $1-P_e$, respectively, the classical Fisher information is defined as
\begin{equation}
\begin{aligned}
F_p=\frac{1}{P_e}(\frac{dP_e}{dp})^2+\frac{1}{P_g}(\frac{dP_g}{dp})^2.
\label{equation9}
\end{aligned}
\end{equation}
For the qubit probe, $P_e+P_g=1$. The $F_p$ can be re-written  as
\begin{equation}
\begin{aligned}
F_p=\frac{1}{P_e(1-P_e)}(\frac{dP_e}{dp})^2.
\label{equation10}
\end{aligned}
\end{equation}
The Fisher information $F_p$  quantifies the sensitivity of the probabilities to infinitesimal changes in the inactive fraction $p$. It bounds the variance of any unbiased estimator $\hat p$ through the Cram\'{e}r--Rao inequality, $\mathrm{var}(\hat p)\ge 1/(\nu F_p)$, where $\nu$ is the number of independent measurements \cite{Cram1946}.

We sweep the inactive fraction $p$ from zero and record the population $P_e$ of the qubit excited-state at each value of $p$. Figure~\ref{aqbpic23}(a) shows the exact numerical results obtained from Eq.~(\ref{equation5}). To characterize the response before $P_e$ vanishes, we fit the curves in the finite-population regime, highlighted by the green region. The fitted curves are expressed as polynomials in $p$,
\begin{equation}
\begin{aligned}
P_e^{(m)}(p)&=\sum_{j=0}^{N_m}x_j^{(m)}p^{j},
\label{equation11}
\end{aligned}
\end{equation}
where \(P_e^{(m)}(p)\) represents the dependence of the excitation numbers on $p$ when the qubit interacts with $m$ active oscillators. When the qubit interacts with $m=1$, $2$, and $3$ active quantum oscillators, the corresponding fitting parameters $N_m$ and $x_j^{(m)}$ are listed in Table~\ref{tab:fit}. The fitted curves show excellent agreement with the numerical results.
\begin{table}[tb]
\caption{Fitting parameters for different numbers of active quantum oscillators.}
\label{tab:fit}
\centering
\begin{tabular}{ccc}
\hline\hline
$m$ & $N_m$ & $\{x_j^{(m)}\} (j=0,1,2,\ldots)$ \\
\hline
1 & 4 &
$(0.0632,\,-0.106,\,-0.0534,\,-0.0173,\,-0.0127)$ \\
2 & 5 &
$(0.181,\,-0.227,\,-0.207,\,-0.264,\,0.101,\,-0.538)$ \\
3 & 6 &
$(0.280,\,-0.243,\,-0.396,\,0.263,\,-4.37,\,9.26,\,-11.8)$ \\
\hline\hline
\end{tabular}
\end{table}

Using the analytical fit in Eq.~(\ref{equation11}), we evaluate the Fisher information with respect to $p$ from Eq.~(\ref{equation10}). Figure~\ref{aqbpic23}(b) shows that the Fisher information develops a pronounced peak as the inactive fraction approaches the aging transition point. This enhancement reflects the critical sensitivity of the probe for $p<p_c$, where the mean-field equations remain nonlinear and the qubit retains a finite excited-state population $P_e$ [Fig.~\ref{aqbpic23}(a)]. Near $p_c$, small changes in $p$ induce a strong variation in $P_e$, producing a sharp increase in $F_p$. For a probe coupled to a single active oscillator, for instance, $F_p$ increases from $32.96$ to $449$ as $p$ changes from $0.46$ to $0.465$, with $F_{\rm max}=449$.

Beyond the aging transition point, the density matrix loses all off-diagonal elements, signaling the absence of coherence among the oscillators \cite{Zhang2025}. Consequently, $M=M_A=M_I=M_{B_+}=0$, and the mean-field master equations become linear. In this regime, Eq.~(\ref{equation5}) gives $\dot{\rho}_a^{\text{ss}}=\beta\mathcal{D}[\sigma_{-}](\rho_a^{\text{ss}})=0,$  so that the probe relaxes to its ground state and $P_e$, and hence $F_p$, vanishes. The sensing performance can be further improved by increasing the number of oscillators coupled to the qubit. As shown in Fig.~\ref{aqbpic23}(b), coupling the probe to three active oscillators enhances the maximum Fisher information by nearly tenfold compared with coupling to fewer oscillators. The number of probe-coupled oscillators must nevertheless remain small, so as not to shift the intrinsic aging transition point of the oscillator network appreciably. These results demonstrate that the Fisher information provides an effective signature of the aging transition. In particular, the sharp enhancement of $F_p$ in a narrow region below the critical point enables high-precision estimation of $p_c$.

\section{Detection of the aging transition point in the classical regime}\label{sec3}

\begin{figure*}[t]
\centering
\includegraphics[width=8.6cm]{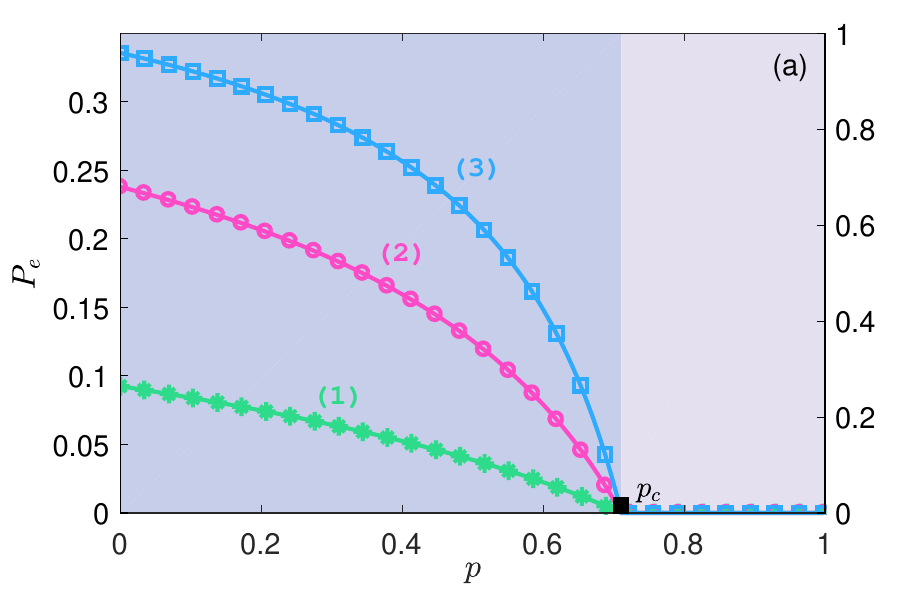}
\includegraphics[width=8.6cm]{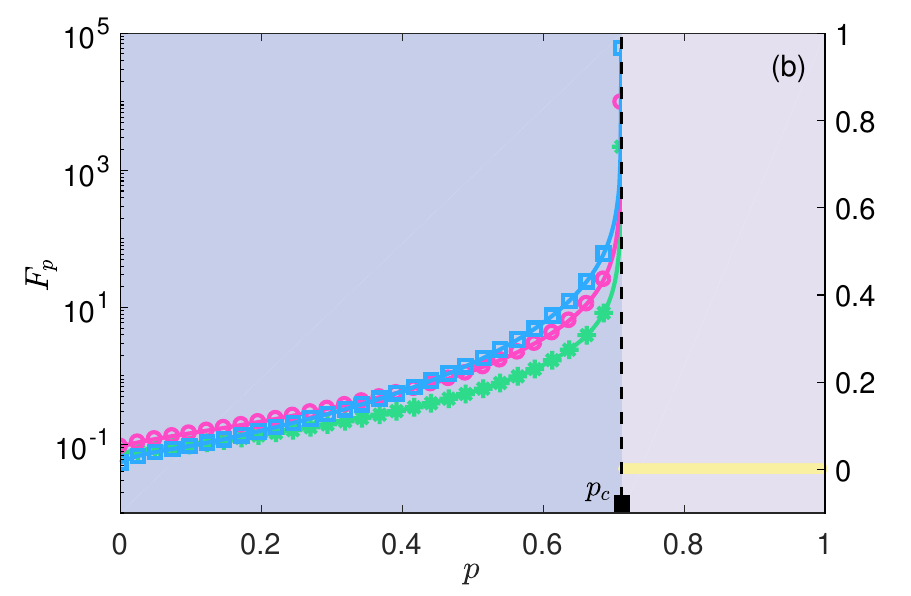}
\caption{Performance of the sensing scheme in the classical regime.
(a) Qubit $|1\rangle$-state population as a function of $p$ for three coupling configurations. The green, pink, and blue curves correspond to the qubit coupled to one, two, and three active classical oscillators, respectively.
(b) Fisher information $F_p$ for estimating the fraction $p$ of inactive oscillators. The part of the curve to the left of the dashed line is plotted on a logarithmic scale, whereas the part to the right is plotted on a linear scale. We set $\gamma_+=80\kappa$ for active oscillators and $\gamma_-=40\kappa$ for inactive oscillators. Other parameters are $N=1000$, $\kappa=1$, $g=8\kappa$, $\beta=300\kappa$, and $V=300\kappa$.}
\label{aqbpic45}
\end{figure*}
Having established the sensing enhancement in the quantum regime, we now examine whether this effect is intrinsically quantum or can persist in the classical limit of large oscillator occupations.

In the regime $\kappa\ll\gamma_\pm$, nonlinear damping is weak compared with the linear gain and loss rates, allowing the oscillators to acquire large occupations. The oscillator operators can then be approximated by classical complex amplitudes, $a_k\rightarrow\alpha_k$. We ask whether the critical enhancement of sensing survives in this classical limit. Starting from the master equation~(\ref{master equation}), the equations of motion for the qubit expectation values follow from $\dot{\langle\mathcal O\rangle}=\mathrm{Tr}(\dot{\rho}\mathcal O)$,
\begin{equation}
\begin{aligned}
\frac{d\langle \sigma_- \rangle}{dt}&=-ig\sum_{k\in S}(\langle a_k\rangle-2\langle a_k\sigma_{+}\sigma_{-}\rangle)-\frac{\beta}{2}\langle\sigma_{-}\rangle,\\
\frac{d\langle \sigma_{+}\sigma_{-} \rangle}{dt}&=-ig\sum_{k\in S}(\langle a_k\sigma_{+}\rangle-\langle a_{k}^{\dag}\sigma_{-}\rangle)-\beta\langle\sigma_{+}\sigma_{-}\rangle.
\label{equation13}
\end{aligned}
\end{equation}
The equations of motion for the oscillators that interact with the qubit are given by
\begin{equation}
\begin{aligned}
\frac{d\langle a_k\rangle}{dt}&=-ig\langle\sigma_{-}\rangle\pm\frac{\gamma_\pm}{2}\langle a_k\rangle-\kappa\langle a_{k}^\dag a_{k}^2\rangle
\\&+\frac{V}{N}\sum_{j=1}^N(\langle a_j\rangle-\langle a_k\rangle),
\label{equation14}
\end{aligned}
\end{equation}
where the equations of motion for the oscillators that do not interact with the qubit are given by
\begin{equation}
\begin{aligned}
\frac{d\langle a_k\rangle}{dt}=\pm\frac{\gamma_\pm}{2}\langle a_k\rangle-\kappa\langle a_{k}^\dag a_{k}^2\rangle+\frac{V}{N}\sum_{j=1}^N(\langle a_j\rangle-\langle a_k\rangle),
\label{equation15}
\end{aligned}
\end{equation}
the sign in front of $\frac{\gamma_\pm}{2}$ is plus for active and minus for inactive oscillators in Eq.~(\ref{equation14}) and Eq.~(\ref{equation15}). In the classical limit regime, we replace the annihilation operator $a_k$ with a complex amplitude $\alpha_k$. Consequently, the following factorization holds:  $\langle a_{k}^\dag a_{k}^2\rangle\rightarrow|\alpha_k|^2\alpha_k$, $\langle a_{k}\sigma_+\sigma_-\rangle\rightarrow\alpha_{k}\langle\sigma_+\sigma_-\rangle$, and $\langle a_{k}\sigma_+\rangle\rightarrow\alpha_{k}\langle\sigma_+\rangle$. For the oscillators, we assume that within each group all oscillators are in the same state. Specifically, for the active oscillators that do not interact with the qubit, we set $\alpha_k = A$; for the inactive oscillators that do not interact with the qubit, we set $\alpha_k = I$; for the active oscillators that interact with the qubit, we set $\alpha_k = \mathcal{A}$ and  for the inactive oscillators that interact with the qubit, we set $\alpha_k = \mathcal{I}$.

Suppose there are $m$ active and $n$ inactive classical oscillators interacting with the qubit. Then we obtain the following equations:
\begin{equation}
\begin{aligned}
\dot{\boldsymbol{X}} = \boldsymbol{M} \boldsymbol{X},
\label{equation16}
\end{aligned}
\end{equation}
where the column vector $\boldsymbol{X}$ is defined as
$
\boldsymbol{X} =
\big(\,\langle\sigma_{-}\rangle,\;
\langle\sigma_{+}\rangle,\;
\langle\sigma_{+}\sigma_{-}\rangle,\;
A,\;
A^*,\;
I,\;
I^*,\;
\mathcal{A},\;
\mathcal{A^*},\;
\mathcal{I},\;
\mathcal{I^*}\,\big)^\mathrm{T},
$
and the specific form of the nonlinear matrix $\boldsymbol{M}$ are given in Appendix~\ref{appendixB}.

Figure~\ref{aqbpic45}(a) shows the qubit excited-state population $P_e$ obtained from Eq.~(\ref{equation16}) in the classical regime. As in the quantum case, we fit the curves in the finite-population regime before $P_e$ vanishes, using the polynomial form in Eq.~(\ref{equation11}). For a probe coupled to $m=1,2,3$ active classical oscillators, the corresponding polynomial orders are $N_1=5$, $N_2=6$, and $N_3=8$, respectively. The results are listed in Table~\ref{tab:classical_fit}.

\begin{table}[H]
\caption{Fitting parameters for different numbers of active classical oscillators.}
\label{tab:classical_fit}
\centering
\begin{tabular}{ccc}
\hline\hline
$m$ & $N_m$ & $\{x_j^{(m)}\} (j=0,1,2,\ldots)$ \\
\hline
1 & 5 &
$(0.093,\,0.081,\,-0.027,\,-0.075,\,0.100,\,-0.011)$ \\
2 & 6 &
$(0.240,\,-0.130,\,-0.190,\,0.510,\,-2.000,\,3.100,\,-2.100)$ \\
3 & 8 &
\parbox[t]{0.58\columnwidth}{
$(0.3357,\,-0.1101,\,-0.3298,\,2.489,\,-16.33,$\\
$55.68,\,-108.5,\,111.2,\,-47.86)$
} \\
\hline\hline
\end{tabular}
\end{table}

Substituting the fitted functions into Eq.~(\ref{equation10}), we obtain the Fisher information as a function of the inactive fraction $p$, as shown in Fig.~\ref{aqbpic45}(b). For classical oscillators, the aging transition is signaled by the vanishing of all oscillator amplitudes $\alpha_k$ \cite{Daido2004,Zhang2025}. Equation~(\ref{equation16}) then shows that, in the steady state, the qubit excited-state population $\langle\sigma_+\sigma_-\rangle=P_e$ also vanishes. Similar to the quantum regime, the outcome probabilities become highly sensitive to small variations in $p$ within a narrow region below the aging transition point, leading to a strong enhancement of the Fisher information. The maximum of $F_p$ therefore indicates that the oscillator network is approaching the onset of the aging transition. As shown in Fig.~\ref{aqbpic45}(b), when the probe is coupled to three classical active oscillators ($p$ correspondingly ranges from 0 to $(N-3)/N$), the critical response enhances $F_p$ by a factor exceeding $10^6$ compared with the small-$p$ regime. We emphasize that similar enhancements are also observed when the qubit is coupled to a few inactive oscillators or to a mixed subset of active and inactive oscillators.

\section{Quantum Fisher information}\label{sec4}
\begin{figure}[b]
\centering
\includegraphics[width=0.45\textwidth]{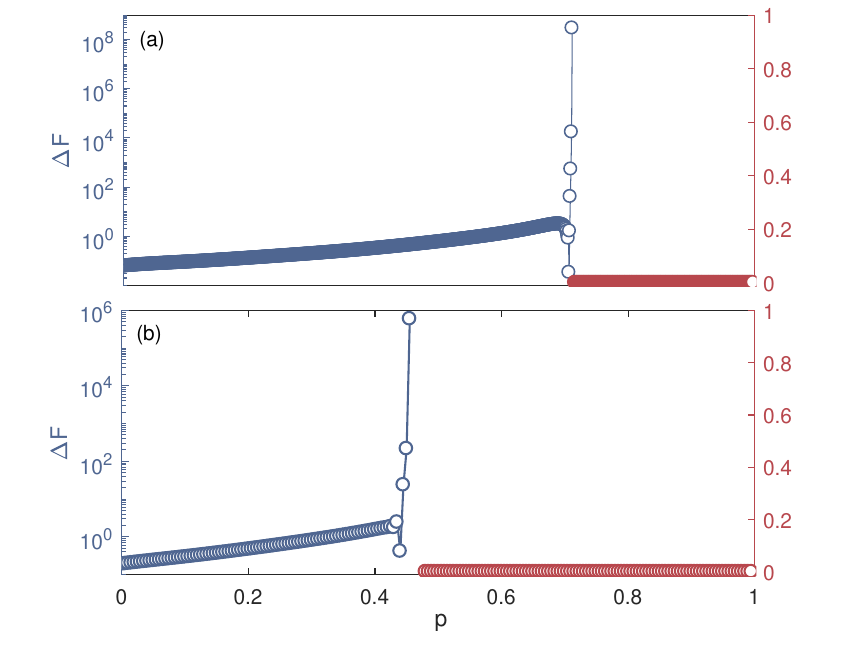}
\caption{Difference $\Delta F$ between the QFI and FI versus $p$ in (a) the classical regime and (b) the quantum regime. The parameters are the same as those in Figs.~\ref{aqbpic45} and \ref{aqbpic23}, respectively. The blue curves are shown on a logarithmic scale, while the red curves are shown on a linear scale.}
\label{aqbpic67}
\end{figure}
The analysis till now has  focused only on the classical Fisher information (FI), whereas the ultimate achievable estimation precision is bounded by the quantum Fisher information (QFI). To show the difference between the two information, we start with  a single-qubit mixed state. The QFI for the qubit in Bloch representation is given by \cite{Zhong2013,Liu2020}
\begin{equation}
\begin{aligned}
\mathcal{F}_a
=|\partial_a\bm{r}|^2
+\frac{\left(\bm{r}\cdot\partial_a\bm{r}\right)^2}
{1-|\bm{r}|^2},
\label{equation17}
\end{aligned}
\end{equation}
where $\bm{r}$ denotes the Bloch vector and $|\bm{r}|$ is the norm of $\bm{r}$. $a$ represents the parameter to be estimated. Here $a$ corresponds to the inactive fraction $p$.  For a single-qubit pure state, this reduces to $\mathcal F_a=|\partial_a\bm{r}|^2$. Calculating the purity of the qubit state  $\text{Tr}\rho_a^2$, we find that in both the quantum and classical regimes, the steady state of $\rho_a$ is a mixed state as long as $P_e$ is not zero. With this knowledge, we now examine the  difference $\Delta F$ between the QFI and the FI as $p$ changes. Consider the case where the qubit couples to three active oscillators, Fig.~\ref{aqbpic67} shows that the QFI grows by several orders of magnitude compared with the FI when $p$ approaches the critical transition point. This observation is physically reasonable, given that the QFI quantifies the maximum attainable Fisher information over all feasible quantum measurements. Qualitatively, however, the QFI and FI exhibit the same dependence on $p$: both quantities feature drastically enhanced sensitivity to $p$ as the parameter approaches the critical value $p_c$. Our findings therefore imply that identification of the transition point can be realized without performing optimization over all feasible measurements.

\section{conclusions}\label{sec5}
In summary, we have proposed and analyzed a quantum sensing scheme for detecting the aging transition in a globally coupled quantum-oscillator network, using a single qubit as a local probe. The qubit is resonantly coupled to a small subset of oscillators, which may be active, inactive, or mixed, and its excited-state population $P_e$ serves as the sensing signal. As the inactive fraction $p$ approaches the aging transition point $p_c$, $P_e$ becomes increasingly sensitive to variations in $p$, giving rise to a sharp enhancement of the Fisher information $F_p$. Once $p$ exceeds $p_c$, the oscillator network loses coherence, the mean-field response vanishes, and the probe relaxes to its ground state with $P_e=0$. The resulting critical enhancement enables high-precision estimation of $p_c$ using only a few probe-oscillator couplings. Increasing the number of coupled oscillators, for example from one to three, further improves the sensing performance; however, this number must remain small to avoid shifting the intrinsic aging transition point. Remarkably, the same critical sensing enhancement persists in the classical limit, where the oscillators are described by complex amplitudes and the Fisher information is dramatically amplified near the transition. These results demonstrate that a simple qubit probe can efficiently estimate aging transition points in both quantum and classical oscillator networks, offering a practical route to detecting critical degradation in dissipative many-body systems.

High-precision detection of the aging transition point is crucial for networks composed of active and inactive oscillators. Such networks arise in a broad range of physical and biological systems, where oscillatory and quiescent units coexist and collectively determine the system dynamics. In these systems, the pronounced enhancement of the Fisher information near the aging transition provides an early-warning signature of the impending collapse of collective oscillatory activity.

\section*{ACKNOWLEDGMENTS}\label{sec5}
This work is supported by the Science Challenge Project (No. TZ2025017) and the National Natural Science Foundation of China (Grant No. 12575010).

\appendix
\section{ $H_x$, $\mathcal{L}_x$ and the  mean-field results in Eq.~(\ref{equation5}) }\label{appendixA}
Here we write the details of $H_x$ and $\mathcal{L}_x$ in Eq.~(\ref{equation5}),
\begin{widetext}
\begin{equation}
\begin{aligned}
H_a &= g(M\sigma_{+}+M^*\sigma_{-}),
&
\mathcal{L}_a &= \beta\mathcal{D}[\sigma_{-}](\rho_a),\\
H_A &= i\frac{V}{N}(M_A a^\dag-{M_A^*}a),
&
\mathcal{L}_A &= \gamma_{+}\mathcal{D}[a^{\dag}](\rho_A)
+\kappa\mathcal{D}[a^2](\rho_A)
+\frac{2V(N-1)}{N}\mathcal{D}[a](\rho_A),\\
H_I &= i\frac{V}{N}(M_I a^\dag-{M_I^*}a),
&
\mathcal{L}_I &= \gamma_{-}\mathcal{D}[a](\rho_I)
+\kappa\mathcal{D}[a^2](\rho_I)
+\frac{2V(N-1)}{N}\mathcal{D}[a](\rho_I),\\
H_{B_\pm} &= g(\langle\sigma_{+}\rangle a+\langle\sigma_{-}\rangle a^{\dag})
+i\frac{V}{N}(M_{B_{\pm}}a^\dag-M_{B_{\pm}}^*a),
&
\mathcal{L}_{B_\pm} &= \gamma_{\pm}\mathcal{D}[\hat{O}](\rho_{B_{\pm}})
+\kappa\mathcal{D}[a^2](\rho_{B_{\pm}})
+\frac{2V(N-1)}{N}\mathcal{D}[a](\rho_{B_{\pm}})
\end{aligned}
\label{A1}
\end{equation}
\end{widetext}
We set \( S_x = \sum_{j\in x} \Tr(a\rho_x) \) $(x=A, I, B_+, B_-)$. The specific forms of $M$ and $M_x$  in Eq.~(\ref{A1}) are given by
\begin{equation}
\begin{aligned}
M       &= S_{B_+} + S_{B_-},  \\
M_A     &= S'_{A}  + S_I + S_{B_+} + S_{B_-},\\
M_I     &= S_A     + S'_I + S_{B_+} + S_{B_-}, \\
M_{B_+} &= S_A     + S_I + S'_{B_+} + S_{B_-}, \\
M_{B_-} &= S_A     + S_I + S_{B_+} + S'_{B_-}.
\label{equationA1}
\end{aligned}
\end{equation}
where $S'_x$ represents $\sum_{j\in x}{}^{'}\text{Tr}(a\rho_x)$.

\section{The specific form of the nonlinear matrix in Eq.~(\ref{equation16})}\label{appendixB}
The nonlinear matrix $\boldsymbol{M}$ in Eq.~(\ref{equation16}) can be written as
\vspace{2cm}
\begin{widetext}
\[
\footnotesize
\boldsymbol{M} =
\left(
\setlength{\arraycolsep}{5pt}
\begin{array}{*{11}{c}}  % 11 ÁоÓÖжÔÆë
-\beta/2 & 0 & 2igC_1 & 0 & 0 & 0 & 0 & -igm & 0 & -ign & 0 \\[4pt]
0 & -\beta/2 & -2igC_1^* & 0 & 0 & 0 & 0 & 0 & igm & 0 & ign \\[4pt]
igC_1^*& -igC_1 & -\beta & 0 & 0 & 0 & 0 & 0 & 0 & 0 & 0 \\[4pt]
0 & 0 & 0 & C_2 & 0 & C_7 & 0 & Vm/N & 0 & Vn/N & 0 \\[4pt]
0 & 0 & 0 & 0 & C_2 & 0 & C_7 & 0 & Vm/N & 0 & Vn/N \\[4pt]
0 & 0 & 0 & C_6 & 0 & C_3 & 0 & Vm/N & 0 & Vn/N & 0 \\[4pt]
0 & 0 & 0 & 0 & C_6 & 0 & C_3 & 0 & Vm/N & 0 & Vn/N \\[4pt]
-ig & 0 & 0 & C_6 & 0 & C_7 & 0 & C_4 & 0 & Vn/N & 0 \\[4pt]
0 & ig & 0 & 0 & C_6 & 0 & C_7 & 0 & C_4 & 0 & Vn/N \\[4pt]
-ig & 0 & 0 & C_6 & 0 & C_7 & 0 & Vm/N & 0 & C_5 & 0 \\[4pt]
0 & ig & 0 & 0 & C_6 & 0 & C_7 & 0 & Vm/N & 0 & C_5
\end{array}
\right),
\]
\end{widetext}
with
$C_1=m\mathcal{A}+n\mathcal{I},
C_2=\frac{\gamma_+}{2}-\frac{V}{N}(Np+m)-\kappa|A|^2,
C_3=-\frac{\gamma_-}{2}-\frac{V}{N}(Nq+n)-\kappa|I|^2,
C_4=\frac{\gamma_+}{2}+\frac{V}{N}(m-N)-\kappa|\mathcal{A}|^2,
C_5=-\frac{\gamma_-}{2}+\frac{V}{N}(n-N)-\kappa|\mathcal{I}|^2,
C_6=\frac{V}{N}(Nq-m),
C_7=\frac{V}{N}(Np-n),$ and $q\equiv1-p$.

\end{document}